\documentclass[apj]{emulateapj}
\journalinfo{Accepted for publication in ApJ}
\submitted{10 December 2008}

\newcommand{\myemail}{hennemann@mpia.de}

\shorttitle{Star-forming cores in a massive clump}
\shortauthors{Hennemann et al.}

\begin{document}

\title{Star-forming cores embedded in a massive cold clump:\\
Fragmentation, collapse and energetic outflows}

\author{Martin Hennemann, Stephan M. Birkmann, Oliver Krause, Dietrich Lemke,\\
Yaroslav Pavlyuchenkov\altaffilmark{1}, Surhud More and Thomas Henning}
\affil{Max-Planck-Institut f\"ur Astronomie, K\"onigstuhl 17, 69117 Heidelberg,
Germany}
\altaffiltext{1}{Current address: Institute of Astronomy of the Russian Academy
of Sciences, Moscow, Russia}
\email{\myemail}

\begin{abstract}
The fate of massive cold clumps, their internal structure and collapse need to
be characterised
to understand the initial conditions for the formation of high-mass stars,
stellar systems, and the origin of associations and clusters.
We explore the onset of star formation in the 75\,M$_\sun$ SMM1
clump in the region ISOSS J18364-0221 using infrared and (sub-)millimetre
observations including interferometry.
This contracting clump has fragmented into two compact cores SMM1 North and
South of 0.05\,pc radius, having masses of 15 and 10\,M$_\sun$, and
luminosities of 20\,L$_\sun$ and 180\,L$_\sun$.
SMM1 South harbours a source traced at 24 and 70\,$\micron$, drives an
energetic molecular outflow, and appears supersonically turbulent at
the core centre.
SMM1 North has no infrared counterparts and shows lower levels of turbulence,
but also drives an outflow.
Both outflows appear collimated and parsec-scale near-infrared features
probably trace the outflow-powering jets.
We derived mass outflow rates of at least
$4\times10^{-5}$\,M$_\sun$\,yr$^{-1}$ and outflow timescales of less than
10$^4$\,yr.
Our HCN(1-0) modelling for SMM1 South yielded an infall velocity of
0.14\,km\,s$^{-1}$ and an estimated mass infall rate of
$3\times10^{-5}$\,M$_\sun$\,yr$^{-1}$.
Both cores may harbour seeds of intermediate- or high-mass stars.
We compare the derived core properties with recent simulations of massive core
collapse. They are consistent with the very early stages dominated
by accretion luminosity.
\end{abstract}


\section{Introduction}

\subsection{Star formation in massive cold clumps}

The early phases of star formation occur in the densest and coldest regions
within molecular clouds, i.e.\ dense cold cores.
Detailed studies of nearby star-forming molecular clouds have led to a
conception of the evolution of individual low-mass stars from prestellar cores
to pre-main-sequence stars
\citep[e.g.][]{1994ApJ...420..837A,2007prpl.conf...17D,2007prpl.conf...33W}.
The formation of high-mass stars (M $\gtrsim 10$\,M$_\sun$) is less understood, in
particular its earliest stages and the link to the origin of associations and
clusters \citep{2007ARA&A..45..481Z,2008ASPC..387.....B}.
Previous studies, mostly aimed towards luminous infrared sources, have provided
evidence for collimated outflows
\citep[recently e.g.][and ref.\ therein]{2007A&A...470..269Z,2005ccsf.conf..105B}
and large-scale rotating structures around forming
high-mass objects \citep[][and ref.\ therein]{2005IAUS..227..135Z}.
This indicates that they also grow via accretion through a disk and
that the high radiation pressure could be released in outflow cavities
\citep{2005ApJ...618L..33K}.
In the low-mass regime, accretion rates are found to be of the order
10$^{-6}$\,M$_\sun$\,yr$^{-1}$ \citep{1997ApJ...485..703W}.
Much higher values of 10$^{-3}$\,M$_\sun$\,yr$^{-1}$ are expected during the build-up
of high-mass stars \citep{2003ApJ...585..850M,2007ApJ...660..479B} and
corresponding increased outflow energetics have been observed
\citep[e.g.\ see compilation by][and ref.\ therein]{2004A&A...426..503W}.
Recently, simulations of protostars accreting at such high rates predicted
a bloating by several tens of solar radii or more and the
accompanying high luminosity at low surface temperature before they reach the
main sequence \citep{2008ASPC..387..189Y,2008ASPC..387..255H}.
This gives further indication that the formation process differs from
a scaled-up version of the low-mass case.

It is presumed that the onset of high-mass star formation occurs in cores
(diameters of 0.1\,pc and less) within massive cold clumps
(diameters of around 0.5\,pc).
Surveys in the far-infrared and at millimetre wavelengths are well suited to
search for such objects, but so far the earliest, i.e.\ prestellar, stages have
eluded discovery. \citet{2007A&A...476.1243M} derive statistical lifetimes
of less than 10$^3$\,yr for the former, and of 10$^4$\,yr for massive dense cores
that are forming stars but exhibit low infrared emission.
Interferometric observations provide the essential spatial resolution to study
the substructure of clumps, and to infer the properties of embedded cores
further requires a good coverage of their spectral energy distribution.
In this work we present such a detailed study of one clump.
In particular, we investigate the amount of fragments in the clump and their
properties, the arising outflow activity, and the indications for collapse.
The clump is located in the region \object{ISOSS J18364-0221} that has been
identified using the ISOPHOT Serendipity Survey
(ISOSS, \citet{1996A&A...315L..64L,2007A&A...466.1205S}) at 170\,$\micron$ to
establish far-infrared colour temperatures.
\citet{2003PhDT.........3K} and \citet{2004BaltA..13..407K} selected sources
containing a large fraction of cold dust, and subsequent studies revealed a
population of cold clumps \citep{2003A&A...398.1007K,2008A&A...485..753H} and a
collapsing massive core \citep{2007A&A...474..883B}.

\subsection{The ISOSS J18364-0221 star-forming region and the SMM1 clump}

The star-forming region \object{ISOSS J18364-0221}
(R.A.\,$18^h36^m24.7^s$, Decl.\,-02\arcdeg21\arcmin49\arcsec [J2000])
was initially studied by \citet{2006ApJ...637..380B}
and is located at a distance of about 2.2\,kpc.
Derived from the near-infrared extinction over a $15\arcmin\times15\arcmin$
field ($\sim$\,100\,pc$^2$), the cloud complex associated with the star-forming
region comprises a mass M $\approx 3200$\,M$_\sun$.
In the same manner, a lower mass limit M $\ge 460$\,M$_\sun$ was found for
the central region ($\sim$\,1\,pc$^2$),
while the far-infrared and submillimetre measurements
give a luminosity L $\approx 800$\,L$_\sun$, an average dust temperature of
about 15\,K and a mass of $900^{+450}_{-330}$\,M$_\sun$.
This region contains two clumps detected in the submillimetre continuum
named SMM1 and SMM2 that are separated by about $1.5\arcmin$ along the east-west
direction.
The SMM1 clump is subject of this publication. Its effective radius is
about 0.2\,pc, and from the thermal emission
a characteristic dust temperature of $16.5^{+6}_{-3}$\,K and a clump mass of
M$_{\rm SMM1} = 75\pm30$\,M$_\sun$ were found. 
The molecular line observations reported in \citet{2006ApJ...637..380B} show
red-shifted self-absorption that is interpreted as signature of collapse motions.
Furthermore, at least one outflow is present.
The outflow properties resulting from the single-dish
observations, i.e.\ the outflow mass of about 18\,M$_\sun$ and the mass outflow
rate of 10$^{-3}$\,M$_\sun$\,yr$^{-1}$, are comparable to values
derived for outflows from presumed high-mass star precursors.
These results render the SMM1 clump a promising object to study the early
phases of collapse and fragmentation occurring in massive cold cores.

The detailed study of this source we present here made use of recently
collected infrared and
(sub-)millimetre data described in the next section. They reveal two
star-forming cores and molecular outflows (Section~3).
In Section~4, we discuss the core properties in the context of
early-stage star formation and present a comparison of the observed HCN spectra
with simple radiative transfer models. Finally, the results are summarised in
Section~5.

\section{Observations and data reduction}

\subsection{Near-infrared observations}
Imaging observations in the J, H, and Ks band towards \object{ISOSS J18364-0221} have been
obtained and are described in \citet{2006ApJ...637..380B}. Additional near-infrared
images in the H$_2$ $\nu$\,=\,1-0 S(1) line ($\lambda =$\,2.122\,$\micron$) were
taken with the Calar Alto 3.5\,m telescope in October 2005
using the prime-focus wide-field camera Omega-2000 \citep{2003SPIE.4841..343B}.
Omega-2000 features a field of view (FOV) of $15.4\times 15.4\,\mathrm{arcmin^2}$
with a pixel scale of $0.4496\arcsec\,\mathrm{pix^{-1}}$. The exposures
were dithered on source to allow for sky subtraction. The data reduction
was done using IRAF.

\subsection{Mid- and far-infrared observations}
IRAC \citep{2004ApJS..154...10F} imaging in all four photometric bands, MIPS
\citep{2004ApJS..154...25R} imaging at 24\,$\micron$ and 70\,$\micron$ and MIPS
spectral energy distribution (SED) mode observations were undertaken with
\textit{Spitzer} \citep{2004ApJS..154....1W}.
The basic flux calibrated imaging data of the \textit{Spitzer Science Center} (SSC)
pipeline were used for further data reduction and analysis. Cosmetic
corrections and astrometric refinement were performed with the MOPEX software
\citep{2005PASP..117.1113M}, and the final images were combined using scripts
in IRAF. Aperture photometry and PSF fitting was done with the
aperture corrections given in the IRAC data handbook and on the SSC
website\footnote{\url{http://ssc.spitzer.caltech.edu/mips/apercorr/}}. The MIPS SED
mode calibration is based on a spectrum of $\alpha$ Boo
\citep{2005ApJ...631.1170L} and the measured MIPS 70\,$\micron$ fluxes.
The calibration uncertainties are about 2\% (IRAC, \citet{2005PASP..117..978R}),
4\% (MIPS 24, \citet{2007PASP..119..994E}), and 10\% (MIPS 70,
\citet{2007PASP..119.1019G}).
The resulting photometric accuracy is estimated to 5\% (IRAC), 10\% (MIPS 24),
and 20\% (MIPS 70 and SED).

\subsection{Submillimetre observations}
The submillimetre continuum observations with SCUBA at the JCMT are outlined in
\citet{2006ApJ...637..380B}.
In light of the results from the interferometric observations that are described
in the next section,
we reanalysed the jiggle maps. The ORAC-DR
\citep{1999ASPC..172..171J} and SURF \citep{1998ASPC..145..216J} software
were used for data reduction and the photometric calibration based on maps
of Uranus acquired shortly before and after the observations. Further analysis
as described in \citet{2001ApJS..134..115S} used the MIRIAD software
\citep{1995ASPC...77..433S}.
The deviations of the JCMT beam from a single Gaussian have been considered
by using the Uranus maps to construct symmetric beam models and deconvolve the
maps of the target regions. The derived beam sizes are 8.2$\arcsec$ at
450\,$\micron$ and 14.8$\arcsec$ at 850\,$\micron$.
The maps were restored with Gaussian beams of 8$\arcsec$ and 14$\arcsec$,
respectively, and fluxes as well as deconvolved source sizes have been derived
by fitting Gaussian components.
The noise levels (1$\sigma$) in the restored maps are 100\,mJy\,beam$^{-1}$ at
450\,$\micron$ and 23\,mJy\,beam$^{-1}$ at 850\,$\micron$.
The photometric accuracy
obtained is estimated to 30\% at 450\,$\micron$ and 20\% at 850\,$\micron$.
For a large aperture covering the SMM1 clump, the photometric results reproduce
those of \citet{2006ApJ...637..380B} within the uncertainty ranges.

\subsection{Millimetre observations}
\label{sec:mmobs}
We have carried out millimetre observations using the IRAM 30m and Plateau de Bure
Interferometer (PdBI). The molecular lines CO(2-1), HCO$^+$(1-0) and HCN(1-0)
have been observed together with the continuum at 1.3\,mm and 3.4\,mm.
The line frequencies are 230.538\,GHz for CO(2-1) and 89.188526\,GHz for
HCO$^+$(1-0).
The HCN(1-0) transition includes three hyperfine components within 4\,MHz at
88.630416\,GHz (F$_{1\rightarrow1}$), 88.631847\,GHz (F$_{2\rightarrow1}$), and
88.633936\,GHz (F$_{0\rightarrow1}$)\footnote{NIST
Recommended Rest Frequencies for Observed Interstellar Molecular Microwave
Transitions: http://physics.nist.gov/restfreq}.
The PdBI configurations C and D were utilised; D was observed in September
2006 and C in April 2007 with the new generation facility receivers. Spectral
resolutions
of 40\,kHz (3\,mm lines), 160\,kHz (CO(2-1)), 1.25\,MHz (3.4\,mm continuum,
$3\times160$\,MHz bandwidth), and 2.5\,MHz (1.3\,mm continuum, $2\times320$\,MHz
bandwidth) were used. Phase calibrators were 1749-096 and 1741-038, additional
amplitude calibrators were MWC349 and 3C273.
Corresponding short-spacing
observations were accomplished in February 2007 at the IRAM 30m as On-the-fly maps using the
single pixel heterodyne receivers with the VESPA correlator and spectral resolutions
of 80\,kHz (1\,mm lines, 80\,MHz bandwidth) and 20\,kHz (3\,mm lines, 40\,MHz
bandwidth). The data were reduced and calibrated with the
GILDAS\footnote{\url{http://www.iram.fr/IRAMFR/GILDAS}} software.
GILDAS was also used to combine the short-spacing and interferometric
data for the lines.
In the case of CO, we chose to convert the measured visibilities to maps
using a weighting scheme that achieves the highest spatial resolution
(synthesised beam size of $1.85\arcsec\times1.33\arcsec$) at the
expense of sensitivity for extended emission.
Therefore the combined CO map does not recover the complete flux.
Compared to the single-dish spectra, the line wing fluxes in the
combined map are lower by a factor of $5\pm1$.
For HCN, the comparison of the extracted spectra shows that the flux measured
in the single-dish data is reproduced in the combined map.
The synthesised continuum beam sizes are $1.9\arcsec\times1.0\arcsec$ (PA
11.6\arcdeg) at 1.3\,mm and $4.6\arcsec\times3.2\arcsec$ (PA 6.1\arcdeg) at
3.4\,mm. We reach noise levels (1$\sigma$) in the cleaned maps of
0.6\,mJy\,beam$^{-1}$ (1.3\,mm continuum),
0.13\,mJy\,beam$^{-1}$ (3.4\,mm continuum),
0.5\,Jy\,beam$^{-1}$\,km\,s$^{-1}$ (CO(2-1)),
0.07\,Jy\,beam$^{-1}$\,km\,s$^{-1}$ (HCO$^+$(1-0)),
and 0.1\,Jy\,beam$^{-1}$\,km\,s$^{-1}$ (HCN(1-0)).
Assuming a dust temperature of 20\,K, we are sensitive to total masses of about
0.2\,M$_\sun$ (3$\sigma$ in the 1.3\,mm continuum).
The noise level (1$\sigma$) in the individual HCN(1-0) spectra used for the
modelling is 0.1\,K.
The 1.2\,mm continuum was observed with the MAMBO-2 bolometer array in March 2007.
The MOPSIC software was used for data reduction
and the noise level (1$\sigma$) in the resulting map is 10\,mJy\,beam$^{-1}$.
The photometric accuracy is estimated to 20\%.

\section{Results}

\subsection{Multiwavelength maps of SMM1}

\begin{figure*}
\epsscale{1.15}
\plotone{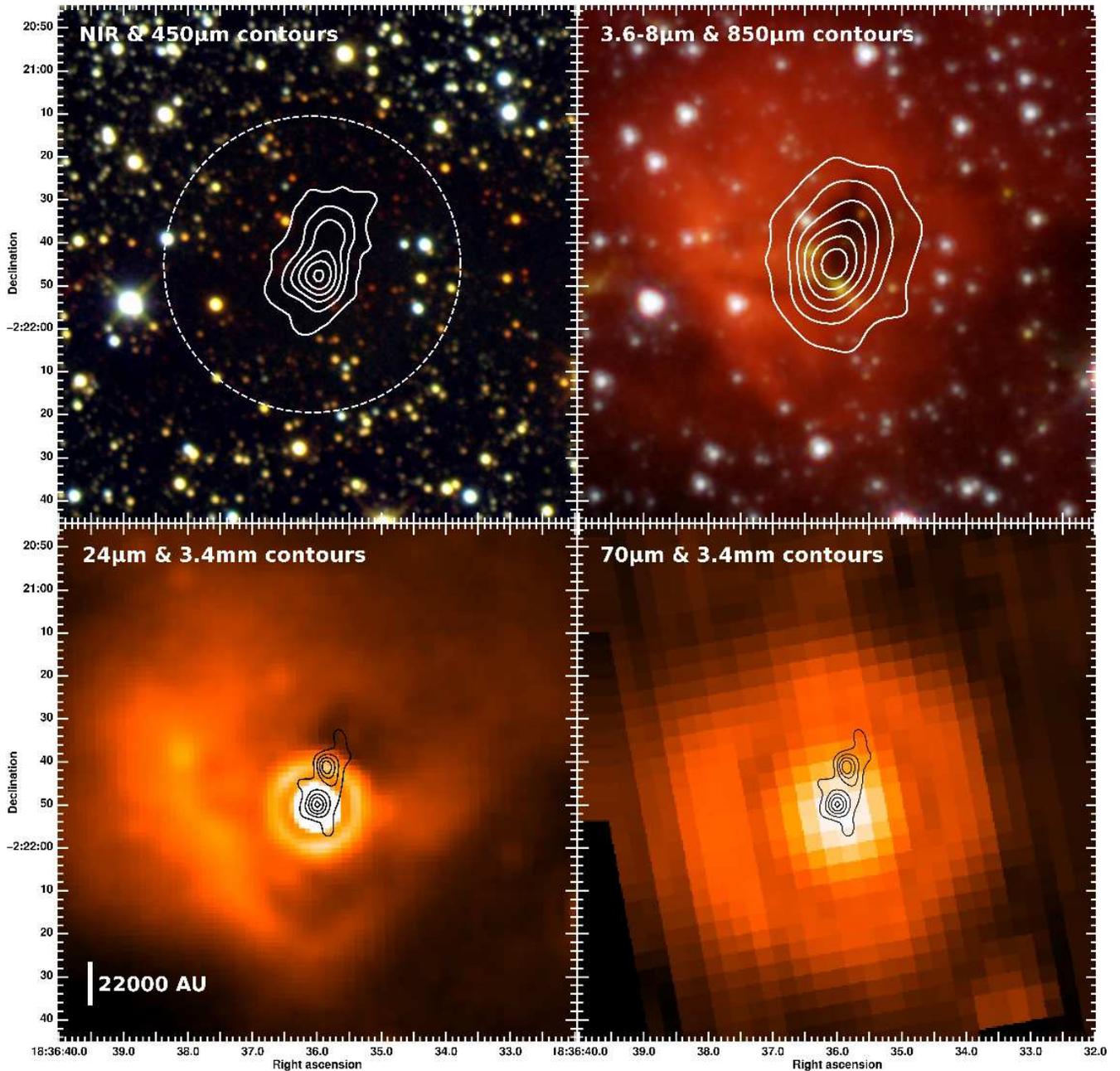}
\caption{Multiwavelength 2$\arcmin\times2\arcmin$ maps towards ISOSS J18364-0221 SMM1.
Upper left: Near-infrared colour composite of J, H, and Ks band observations
with Omega-2000 and solid contours of the SCUBA 450\,$\micron$ map
(5$\sigma$, 10$\sigma$,\,\dots).
The dashed contour shows the region used for background subtraction in the
450 and 850\,$\micron$ maps.
Upper right: Mid-infrared colour composite of the 4 \textit{Spitzer} IRAC channels
and contours of the SCUBA 850\,$\micron$ map (5$\sigma$, 10$\sigma$,\,\dots).
Lower left: \textit{Spitzer} MIPS 24\,$\micron$ map and contours of the PdBI 3.4\,mm
continuum observations (3$\sigma$, 6$\sigma$,\,\dots).
Lower right: \textit{Spitzer} MIPS 70\,$\micron$ map and contours of the PdBI 3.4\,mm
continuum observations (3$\sigma$, 6$\sigma$,\,\dots).
The 3.4\,mm contours in the lower panels reveal the two cores SMM1 North and
South.
}
\label{fig:4map}
\end{figure*}

\begin{figure*}
\epsscale{1.15}
\plotone{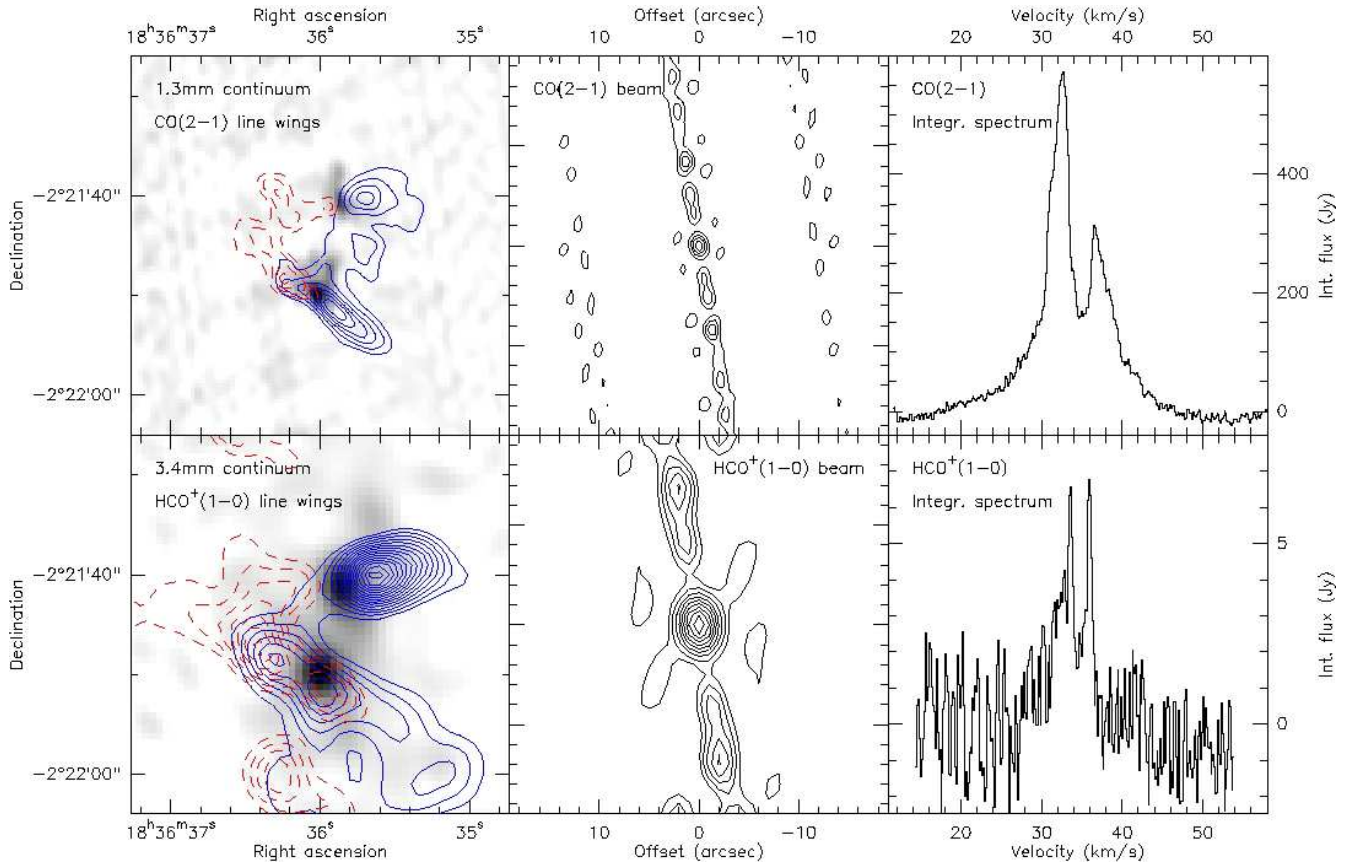}
\caption{ISOSS J18364-0221 SMM1 CO(2-1) and HCO$^+$(1-0) observations.
Top row: Observations at 1.3\,mm; continuum emission is shown in grey-scale,
and CO(2-1) line emission in dashed red (velocity range 20\dots30\,km\,s$^{-1}$;
contours at 2, 3,\,\dots, 7\,Jy\,beam$^{-1}$\,km\,s$^{-1}$)
and solid blue (40\dots50\,km\,s$^{-1}$;
contours at 3, 4,\,\dots, 9\,Jy\,beam$^{-1}$\,km\,s$^{-1}$).
Besides, the beam pattern in steps of 20\% and the
integrated \mbox{CO(2-1)} spectrum are plotted.
Bottom row: Observations at 3.4\,mm; continuum emission is shown in grey-scale,
and HCO$^+$(1-0) line emission in dashed red (velocity range 20\dots30\,km\,s$^{-1}$;
contours at 60, 80,\,\dots, 200\,mJy\,beam$^{-1}$\,km\,s$^{-1}$)
and solid blue (40\dots50\,km\,s$^{-1}$;
contours at 100, 120,\,\dots, 400\,mJy\,beam$^{-1}$\,km\,s$^{-1}$).
Besides, the beam pattern in steps of 10\% and the
integrated HCO$^+$(1-0) spectrum are plotted.
}
\label{fig:pdbimap}
\end{figure*}

In Figs.~\ref{fig:4map} and \ref{fig:pdbimap}
we show multiwavelength maps observed towards \object{ISOSS J18364-0221} SMM1.
In the interferometric millimetre continuum observations SMM1 is resolved into
two cores SMM1 North and SMM1 South (Fig.~\ref{fig:4map} lower panels),
separated by $9.5\arcsec$ (ca.\ 21000\,AU).
In the near-infrared (Fig.~\ref{fig:4map} upper left), noticeable extinction and
reddening is
present in the surrounding of SMM1. There are no near-infrared sources detected
towards the (sub-)millimetre peaks.
The mid-infrared measured with IRAC (Fig.~\ref{fig:4map} upper right) is dominated by extended
emission in
the 8\,$\micron$ band along the outer rim of SMM1. Towards the centre of SMM1,
a filamentary structure is observed in absorption, and no stellar objects are
associated.
At 24\,$\micron$ (Fig.~\ref{fig:4map} lower left) a point source at SMM1 South
shows up. At this
wavelength, the extended emission around SMM1 is also visible, and in the
north-west a dip remains, but the absorption feature is superposed by the PSF
of the southern source.
No obvious 24\,$\micron$ source towards SMM1 North is detected.
Similarly, SMM1 South emits at 70\,$\micron$ (Fig.~\ref{fig:4map} lower right).
The wide PSF overlays
possible extended emission at the rim of SMM1, and no obvious emission at SMM1
North is detected.

In the interferometric 1.3\,mm continuum map (Fig.~\ref{fig:pdbimap} top row),
both SMM1
North and South appear slightly extended to the north, but the beam sidelobes
may affect the morphology. At 3.4\,mm they appear unresolved
(Fig.~\ref{fig:pdbimap} bottom row), and none of the extensions are traced.
The contours in Fig.~\ref{fig:pdbimap} derived from the
emission in the line wings of the CO(2-1) and HCO$^+$(1-0) transitions reveal
two molecular outflows. SMM1 South constitutes the origin of a
north-east-to-south-west (PA ca.\ 50\arcdeg) outflow that also gives rise to the
mid-infrared features. Towards SMM1 North an outflow in east-to-west direction
(PA ca.\ -80\arcdeg) is found.
Its red lobe blends into the red lobe of the SMM1 South outflow.
Most prominent in CO(2-1) is the SMM1 South blue outflow lobe that appears
collimated. In HCO$^+$(1-0), the blue lobe of the SMM1 North outflow
is strongest and it also shows a rather low outflow opening angle, but in both
cases the outflows are traced only close to the cores and we could not
derive an accurate quantitative measure.

\begin{figure*}
\epsscale{1.15}
\plotone{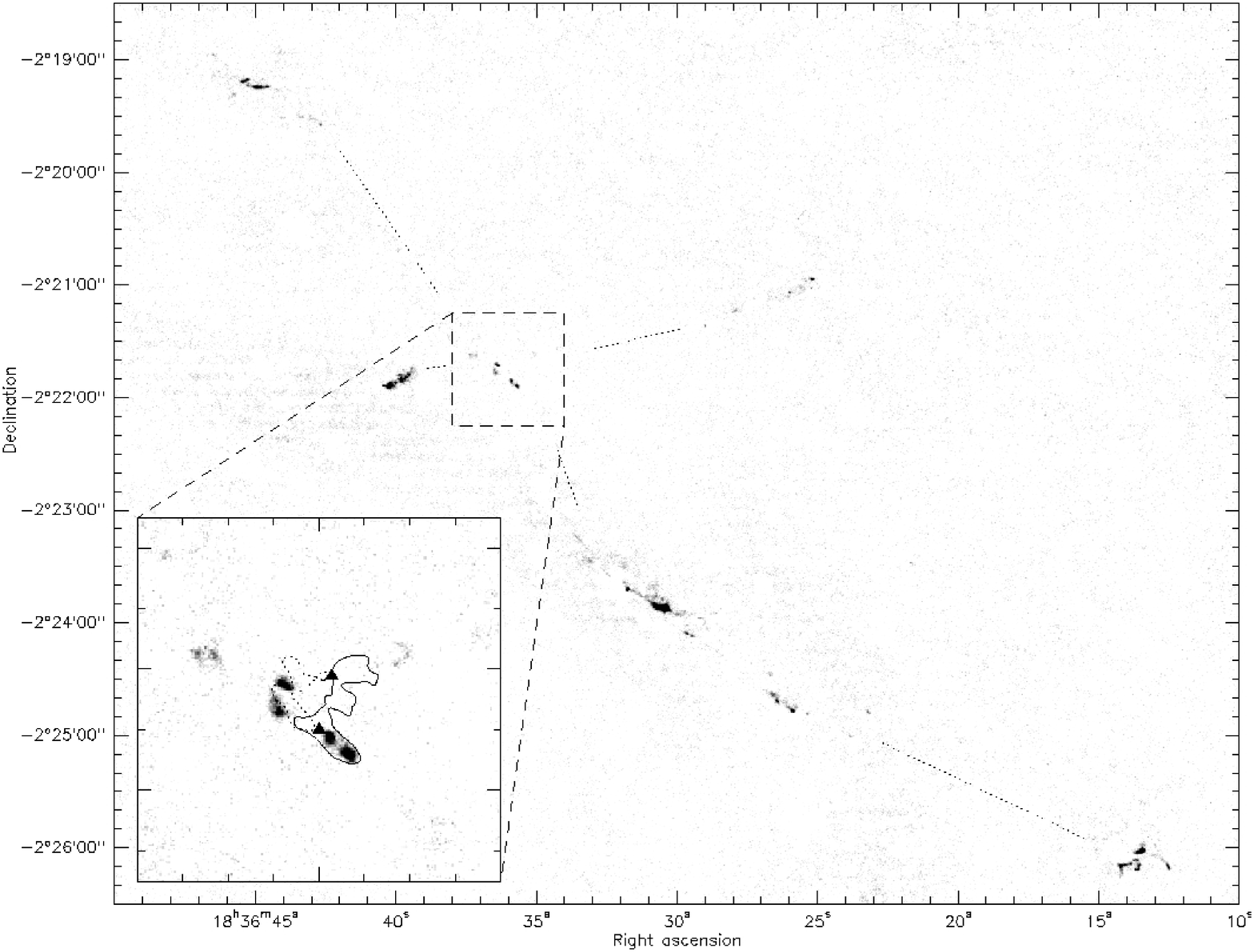}
\caption{H$_2$ ($\lambda =$\,2.122\,$\micron$) emission map towards ISOSS
J18364-0221. The dotted lines indicate the proposed connection of the emission
features to the different outflow lobes.
The inlay shows a magnification of the region around SMM1. The positions
of the SMM1 North and South cores are marked with filled triangles, and the
CO(2-1) line wing emission of the red (dotted contour, 2.5\,Jy\,beam$^{-1}$\,km\,s$^{-1}$) and blue
(solid contour, 3.5\,Jy\,beam$^{-1}$\,km\,s$^{-1}$) molecular
outflow lobes are over-plotted (cf. Fig.~\ref{fig:pdbimap} top row).
}
\label{fig:h2map}
\end{figure*}

In Fig.~\ref{fig:h2map} we show the near-infrared H$_2\ \nu$\,=\,1-0 S(1)
($\lambda =$\,2.122\,$\micron$) line emission map.
This map has been derived by subtracting the scaled K band image from
the narrow-band image so that stars cancelled out. Stars that were not properly
subtracted have been masked afterwards for clarity.
As shown in the inlay of Fig.~\ref{fig:h2map}, towards the blue and red lobes of
the molecular outflow driven by SMM1 South there is patchy near-infrared
H$_2$ emission.
Such features trace protostellar jets, and the emission probably arises
from collisionally excited H$_2$ in shocked gas
\citep[][and ref.\ therein]{1980ApJ...241..728E,2001ARA&A..39..403R}.
This is supported by faint filamentary emission in the 4.5 and 5.8\,$\micron$
bands we detected lateral to SMM1 South along north-east-to-south-west and
in the vicinity of SMM1 North (yellowish features in Fig.~\ref{fig:4map} top right).
These bands contain several H$_2$ lines
\citep{2004ApJS..154..352N,2005MNRAS.357.1370S}.
Only a weak near-infrared feature is detected close to the blue lobe of the SMM1
North outflow. On the large-scale
map we found additional filaments of H$_2$ emission. Though they are not
exactly aligned with the molecular outflows, they can be assigned to individual
lobes which would make it necessary that the proposed jets show bending
resulting in S-shapes. The outermost features in the North-East and South-West
are located at a projected distance of about 6.7\,pc from each other,
and the South-West feature is at a projected distance of about 4.6\,pc from
SMM1 South.
Among the sources detected in the vicinity of the features, we do not
find other candidates for jet-driving young stellar objects.

%\clearpage

%\clearpage

\subsection{SEDs and core sizes}

In the near-infrared and at 3.6\,$\micron$ no emission is coinciding with SMM1
North or South and we derived upper flux limits from our maps.
Close to SMM1 North, the emission in the 4.5 and 5.8\,$\micron$ IRAC bands
were measured on levels of 71 and 155\,$\mu$Jy.
In the 4.5\,$\micron$ band there is extended emission overlapping with the SMM1
South position, where the flux is about 1.2\,mJy.
At 8\,$\micron$ no counterparts corresponding to SMM1 North or South are
detected.

The source towards SMM1 South clearly dominates the emission at 24 and
70\,$\micron$.
Due to the separation of about 9$\arcsec$ between SMM1 North and South,
the derivation of fluxes at these wavelengths is impaired towards SMM1 North
(PSF FWHM of about 6$\arcsec$ and 17$\arcsec$).
We accomplished PSF subtraction to remove the emission of SMM1 South.
At both 24 and 70\,$\micron$ the residual maps do not reveal a second compact
source at SMM1 North. The 3$\sigma$ upper limits for SMM1 North were derived
from the residual scatter at its position. We ended up with 24\,$\micron$ fluxes
of 199\,mJy and 15.7\,Jy for SMM1 South and upper flux limits of 5.3\,mJy and
177\,mJy for SMM1 North at 24 and 70\,$\micron$ respectively.

To derive submillimetre fluxes for the two internal SMM1 components we fitted
a Gaussian to each peak on the 450\,$\micron$ map and an extended Gaussian to
account for the surrounding clump emission. The same sizes convolved with the
14$\arcsec$ Gaussian were then used as fixed parameters to extract 850\,$\micron$
fluxes.
The clump background is well fit with a deconvolved FWHM size of 68$\arcsec$
and total fluxes of 12.3\,Jy (450\,$\micron$) and 2.02\,Jy (850\,$\micron$).
For SMM1 South we got a deconvolved FWHM size of 5.9$\arcsec$ and fluxes of
4.42\,Jy (450\,$\micron$) and 622\,mJy (850\,$\micron$), and for SMM1 North a size
of 7.0$\arcsec$ and fluxes of 2.66\,Jy (450\,$\micron$) and 476\,mJy (850\,$\micron$).
The total fluxes we derived are higher than those given in
\citet{2006ApJ...637..380B} because we allowed for a larger extent of the background
component.
As in the 850\,$\micron$ map (Fig.~\ref{fig:4map}), SMM1 North and
South blend into a single elongated maximum in the single-dish 1.2\,mm map.
Therefore we extracted fluxes in the same way for both sources,
using the positions from the submillimetre and allowing for a pointing offset.
The derived fluxes are 151\,mJy for SMM1 South and 156\,mJy for SMM1 North.

\begin{figure}
\epsscale{1.15}
\plotone{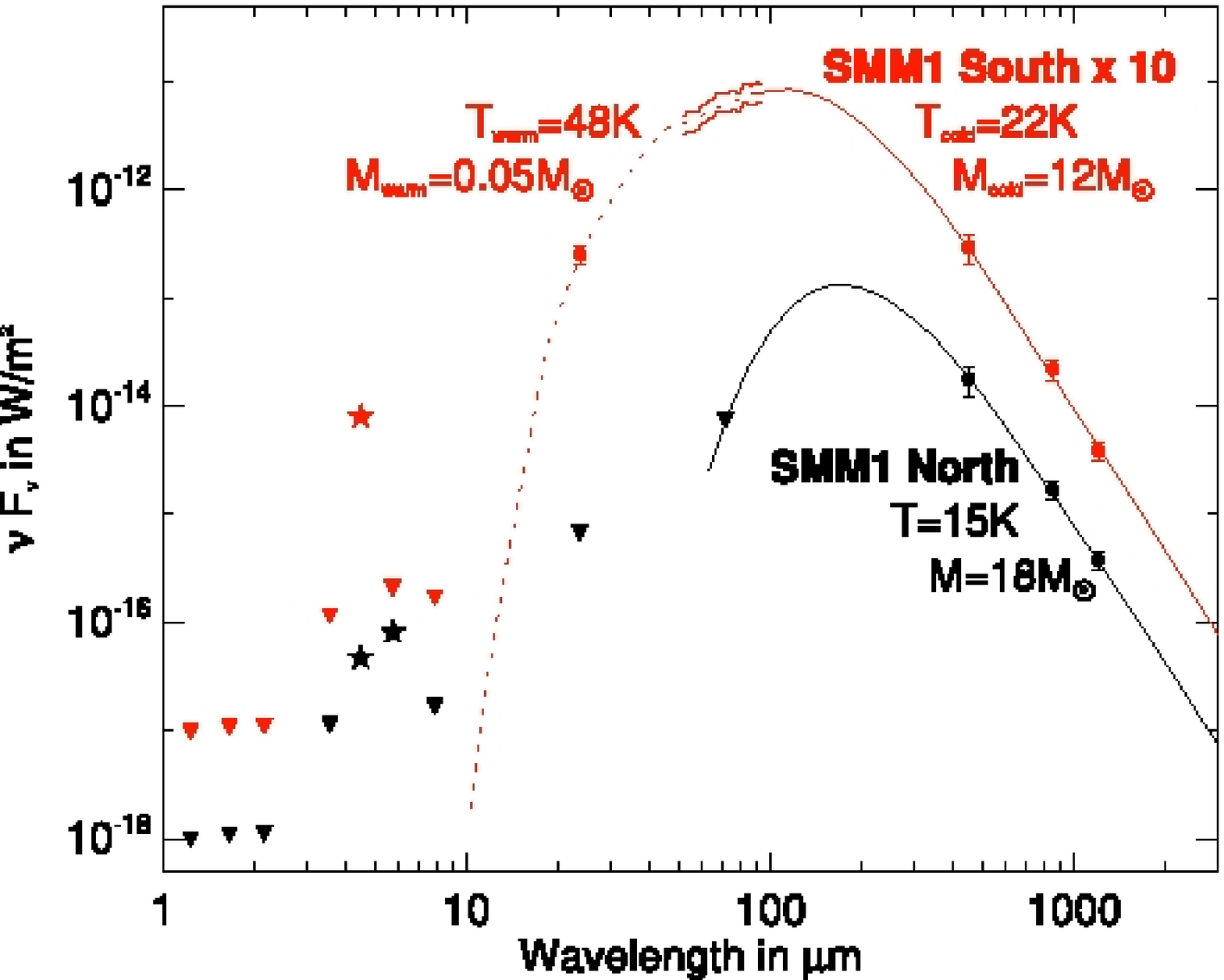}
\caption{Spectral energy distributions of ISOSS J18364-0221 SMM1 North (black)
and South (red, scaled up by 10). Upper limits are indicated by triangles.
The dots represent the data points obtained with SCUBA (850 and 450\,$\micron$)
and MIPS (24\,$\micron$). Histogram-like bars show the error range of the MIPS
SED spectrophotometry (53 to 93\,$\micron$) for SMM1 South.
Stars represent emission
in the IRAC bands (3.6, 4.5, 5.8 and 8.0\,$\micron$) associated with the core
positions that stems from jet features (see text). The J, H and Ks bands were
observed with Omega-2000.
The lines represent fits to the SEDs via modified Planck components
(see text).}
\label{fig:sed_both}
\end{figure}

The continuum fluxes for both cores are listed in Table~\ref{tab:corepars}.
Fig.~\ref{fig:sed_both} shows the spectral energy distributions of the
continuum emission towards the two sources including the jet features in the
IRAC bands. The curves fitted to the data are described in the next subsection.

From the interferometric continuum maps we also derived fluxes for both
sources by fitting Gaussian components. In the case of SMM1 South, 80\,mJy
(1.3\,mm) and 4.1\,mJy (3.4\,mm) and FWHM dimensions of
$4.4\arcsec\times3.6\arcsec$ (PA -50\arcdeg) were measured,
corresponding to a core size of $\sim$9000\,AU.
For SMM1 North the fluxes are 49\,mJy (1.3\,mm) and 3.9\,mJy (3.4\,mm)
and FWHM sizes of $6.3\arcsec\times3.1\arcsec$ (PA 34\arcdeg) were measured,
corresponding to a more elliptical core of $14000\times7000$\,AU.

We also estimated the bolometric luminosities from the two SEDs.
SMM1 South has a luminosity L $\approx 180$\,L$_\sun$, and
for SMM1 North we got L $\approx 20$\,L$_\sun$ using the upper limit fluxes
in the infrared.

%\clearpage
%\clearpage

\subsection{Dust temperatures and masses}

\citet{1994A&A...291..943O} have derived dust opacities for dense protostellar
cores and for our analysis we used the model with thin ice mantles and
coagulation at a gas density of n$_{\rm H} = 10^6$\,cm$^{-3}$ (OH5,
$\kappa_{\rm 1.3mm}$= 0.9\,cm$^2$\,g$^{-1}$).
To estimate the temperature uncertainties we also applied the model without
ice mantles (OH2) and the initial, non-coagulated opacities (OH1).
In Fig.~\ref{fig:sed_both} we show the OH5 fit to the emission of SMM1 South using
two modified Planck components assuming optically thin emission.
The long-wavelength SED is matched by thermal dust emission at
$22^{+4}_{-1}$\,K (solid line).
From the fit we derived a dust mass of M$_{\rm d}^{\rm cold}=0.12$\,M$_\sun$.
The fluxes at shorter wavelengths are reproduced by a dust component at 48\,K
with a dust mass of M$_{\rm d}^{\rm warm}=5\times10^{-2}$\,M$_\sun$ (dashed line).
For the SMM1 North source we used the upper flux limit at 70\,$\micron$ to derive
an upper limit dust temperature of $15^{+1}$\,K from a single component fit also
shown in Fig.~\ref{fig:sed_both}. The corresponding dust mass is
M$_{\rm d}=0.18$\,M$_\sun$. Assuming a canonical gas-to-dust mass ratio of
100, the masses of the cores are ca. 12\,M$_\sun$ (SMM1 South) and
18\,M$_\sun$ (SMM1 North), with an uncertainty of about a factor two
(see~\ref{sec:coreproperties}).

Preserving the dust temperatures and masses, the expected continuum fluxes at
3.4\,mm are 4.9\,mJy (SMM1 South) and 4.8\,mJy (SMM1 North), thus the measured
fluxes trace about 10\,M$_\sun$ for SMM1 South and about 15\,M$_\sun$
for SMM1 North as compact components. The differences stem from the filtering in
the interferometric
observations.
Using these masses and the core FWHM sizes of 9000\,AU and 10000\,AU, we
calculated volume-averaged densities
$\langle$n$_{\rm H_2}\rangle$~=~3M/(4$\pi$\,FWHM$^3$) which are
$4\times10^5$\,cm$^{-3}$ for SMM1 South and $5\times10^5$\,cm$^{-3}$ for SMM1
North.
The remaining mass of the SMM1 clump is located in a more extended envelope.
From the dust temperatures and peak fluxes in the interferometric 1.3\,mm
continuum map we derived peak column densities of $2.7\times10^{23}$\,cm$^{-2}$
(SMM1 North) and $2.4\times10^{23}$\,cm$^{-2}$ (SMM1 South).
Table~\ref{tab:corepars} summarises the properties of the two detected cores.

%\clearpage
\begin{deluxetable*}{lcc}
\tablecaption{Continuum flux measurements and derived core properties.
\label{tab:corepars}}
\tablehead{
\colhead{Parameter}
& \colhead{SMM1 North}
& \colhead{SMM1 South}
}
\startdata
24\,$\micron$ flux (mJy) & $<$5.3 & 199\\
70\,$\micron$ flux (mJy) & $<$177 & 15700\\
450\,$\micron$ flux (mJy) & 2660 & 4420\\
850\,$\micron$ flux (mJy) & 476 & 622\\
30m 1.2\,mm flux (mJy) & 156 & 151\\
PdBI 1.3\,mm flux (mJy) & 49 & 80\\
PdBI 3.4\,mm flux (mJy) & 3.9 & 4.1\\
\hline
FWHM size (AU) & 14000$\times$7000 (PA 34\arcdeg) & 10000$\times$8000 (PA -50\arcdeg)\\
Luminosity (L$_\sun$) & 20 & 180\\
Dust temperature\tablenotemark{a} (K)& 15 & 22\\
N$_{\rm H_2}^{\rm peak}$\tablenotemark{b} (cm$^{-2}$) & $2.7\times10^{23}$ & $2.4\times10^{23}$\\
Mass\tablenotemark{c} (M$_\sun$) & 15 & 10\\
$\langle$n$_{\rm H_2}\rangle$ (cm$^{-3}$) & $5\times10^5$ & $4\times10^5$\\
\enddata
\tablenotetext{a}{Derived from the single-dish continuum flux measurements/upper limits between 70 and 1200\,$\micron$.}
\tablenotetext{b}{Derived from the interferometric 1.3\,mm continuum map and the given dust temperatures.}
\tablenotetext{c}{Derived from the interferometric 3.4\,mm continuum map and the given dust temperatures.}
\end{deluxetable*}
%\clearpage

\subsection{Properties of the molecular outflows}

From our CO(2-1) map we inferred the properties of the outflows.
For this purpose it is necessary to convert the measured CO emission into
molecular hydrogen column densities N$_{\rm H_2}$, and we used the relation
N$_{\rm H_2} = 3\times10^{20}$\,cm$^{-2}$\,K$^{-1}$\,km$^{-1}$\,s\,$\times
\int$T$_{\rm mb}$\,dv from \citet{1997ApJS..110...71O}. The mass was then
calculated as M\,$= \mu$m$_{\rm H_2}$d$^2 \sum$\,N$_{\rm H_2} \Delta \Omega$,
where $\mu$ is the ratio of gas mass to hydrogen mass (taken to be 1.36),
m$_{\rm H_2}$ is the mass of the hydrogen molecule, and $\Delta \Omega$ is the
solid angle covered by N$_{\rm H_2}$. The integrated emission in the
CO map towards SMM1 traces 4.8\,M$_\sun$, compared to a
mass of 75\,$\pm$\,30\,M$_\sun$ derived from the dust continuum emission
\citep{2006ApJ...637..380B}.
This discrepancy probably stems from the uncertainty of the above relation in
combination with the missing flux in the CO map (see~\ref{sec:mmobs}).
Apparent also in the integrated CO(2-1) spectrum in Fig.~\ref{fig:pdbimap},
the optical depth in the line is significant.
Other outflow studies have found CO(2-1) optical depths on the order of 10
referring to a comparison with {}$^{13}$CO emission as described in
\citet{1993ApJ...417..624C}.
In the case of SMM1 South, the fact that near-infrared H$_2$ emission is observed
towards both molecular outflow lobes indicates that the difference in extinction
is not very large. This suggests a high inclination $i$ of the outflow axis
with respect to the line-of-sight.
Because we cannot further constrain the inclination, we assume $i = 57.3\arcdeg$
for both outflows in the following, corresponding to the mean of a random
distribution of outflow orientations.
In Table~\ref{tab:outflowpars} we list the parameters derived as described in
\citet{2000A&A...364..613H} for the different
outflow lobes, corresponding to line emission above 3$\sigma$.
Mean outflow velocities were derived from the mechanical momenta P and the
masses using $<$v$> =$P/M, and halves of the lobe extents
in outflow direction were taken as travelled distance to calculate the dynamical
timescales.
We interpret the given masses as lower limits, and
as described in \ref{sec:mmobs}, the true masses are probably higher by roughly a
factor of five.
The same applies to the mechanical momenta and the
kinetic energies.
While the mass derivation may be precise within factors of 2 to 4
\citep[cf.][]{1990ApJ...348..530C}, the dynamical
parameters are less certain and can be considered as order of magnitude estimates.
Besides, the blending of the two red outflow lobes hampered our parameter
derivation, so we regard the values derived for the blue lobes as more precise.

%\clearpage
\begin{deluxetable*}{lcccc}
\tabletypesize{\scriptsize}
\tablecaption{Outflow parameters derived from the CO(2-1) map.
\label{tab:outflowpars}}
\tablehead{
\colhead{Outflow Parameter}
& \colhead{South Red Lobe}
& \colhead{South Blue Lobe}
& \colhead{North Red Lobe}
& \colhead{North Blue Lobe}
}
\startdata
Proj.\,Velocities (km\,s$^{-1}$) & 40\dots50 & 20\dots30 & 40\dots50 & 20\dots30\\
Solid Angle (as$^2$) & 86 & 195 & 27 & 119\\
Mass\tablenotemark{a} (M$_\sun$) & 0.066 & 0.15 & 0.023 & 0.12\\
Mech.\,Momentum\tablenotemark{b} (M$_\sun$\,km\,s$^{-1}$) & 0.85 & -2.3 & 0.30 & -1.8\\
Kin.\,Energy\tablenotemark{b} (J) & 1.7$\times$10$^{38}$ & 2.5$\times$10$^{38}$ & 1.8$\times$10$^{38}$ & 2.4$\times$10$^{38}$\\
Dyn.\,Timescale\tablenotemark{b} (yr) & 7.7$\times$10$^3$ & 7.0$\times$10$^3$ & 3.8$\times$10$^3$ & 5.8$\times$10$^3$\\
Mass Outflow Rate\tablenotemark{b} (M$_\sun$\,yr$^{-1}$) & 8.6$\times$10$^{-6}$ & 2.1$\times$10$^{-5}$ & 6.0$\times$10$^{-6}$ & 2.1$\times$10$^{-5}$\\
Mech.\,Luminosity\tablenotemark{b} (L$_\sun$) & 1.8 & 2.9 & 3.9 & 3.4\\
\enddata
\tablenotetext{a}{Masses are lower limits because of missing flux.}
\tablenotetext{b}{Assuming an inclination $i = 57.3\arcdeg$ of the outflow axes.}
\end{deluxetable*}
%\clearpage

\subsection{HCN emission towards SMM1}

\begin{figure*}
\epsscale{1.15}
\plotone{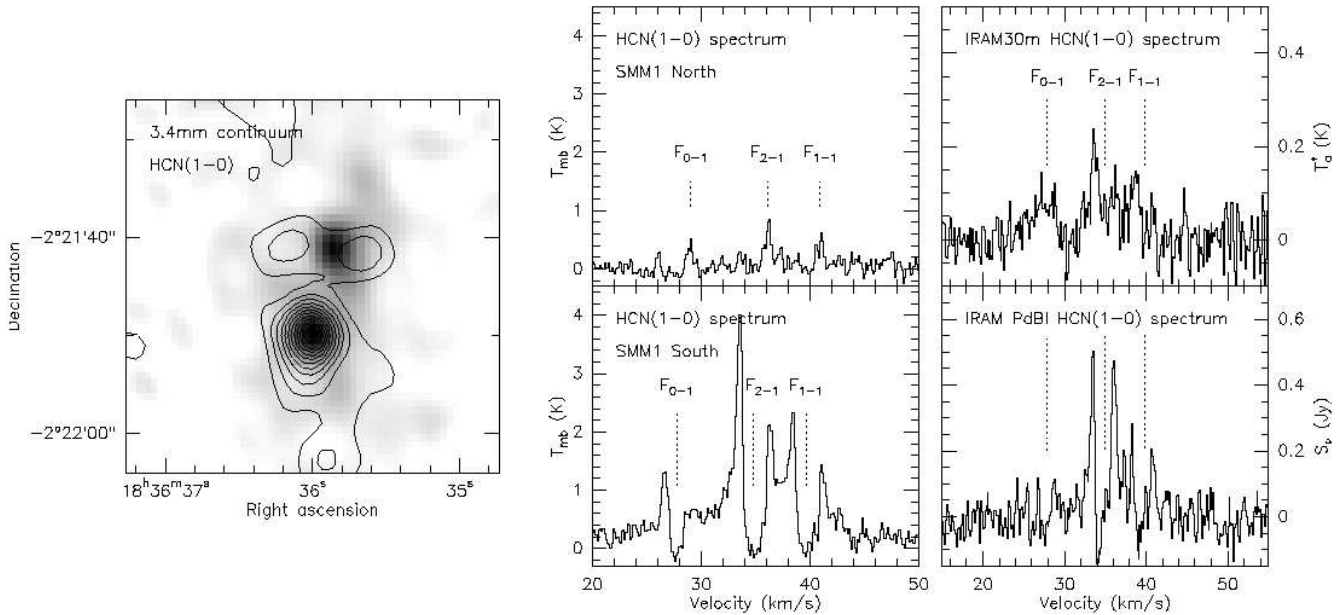}
\caption{ISOSS J18364-0221 SMM1 HCN(1-0) observations.
Left panel: The 3.4\,mm continuum emission is shown in
grey-scale, and the contours give the integrated HCN(1-0) line emission
at 0.1, 0.2,\,\dots, 1.1\,Jy\,beam$^{-1}$\,km\,s$^{-1}$.
The middle panels show the HCN(1-0) spectra towards SMM1 North and South from the
combined IRAM 30m and PdBI data.
The hyperfine components are marked for v$_{\rm LSR}=36.1$\,km\,s$^{-1}$
(SMM1 North) and v$_{\rm LSR}=34.8$\,km\,s$^{-1}$ (SMM1 South).
In the right panels the IRAM 30m single-dish spectrum and the IRAM PdBI
spectrum before combination are plotted. There, the hyperfine
components are marked for v$_{\rm LSR}=35$\,km\,s$^{-1}$.
}
\label{fig:pdbihcn}
\end{figure*}

The three HCN(1-0) hyperfine components F$_{1\rightarrow1}$, F$_{2\rightarrow1}$,
and F$_{0\rightarrow1}$ lie within 4\,MHz and the expected line ratios are
3:5:1 in the optically thin case for local thermal equilibrium (LTE).
The \mbox{HCN(1-0)} map integrated over all three components
(Fig.~\ref{fig:pdbihcn} left) shows a pronounced maximum
towards SMM1 South, whereas only weak emission is detected at SMM1 North.
It is possible that the minor peaks that are offset from SMM1 North stem from
the molecular outflows, because it has been shown recently that HCN can be
present there as well \citep{2007A&A...470..269Z}.
The HCN(1-0) spectra towards the centre positions of SMM1 South and North are
shown in the middle panels of Fig.~\ref{fig:pdbihcn}.

In the case of SMM1 North, the signal-to-noise ratio of HCN(1-0) is rather
low ($\approx$ 6) and the line shapes are uncertain.
As indicated in the plot, the positions of the peaks are consistent with a
systemic velocity v$_{\rm LSR}=36.1$\,km\,s$^{-1}$.
The observed line ratios (neglecting the noise) depart from the expected
optically thin LTE ratios:
F$_{0\rightarrow1}$/F$_{2\rightarrow1}=0.6$ and
F$_{1\rightarrow1}$/F$_{2\rightarrow1}=0.75$ 
compared to 0.2 and 0.6 in the LTE case.
Such hyperfine ``anomalies'' have also been observed
towards star-forming clouds in the past \citep{1982ApJ...258L..75W}.
\citet{1993A&A...279..506G} have investigated this numerically
and found that the velocity structure in the emitting cloud core
could strongly affect the line ratios. In particular, they found
that an increased F$_{1\rightarrow1}$/F$_{2\rightarrow1}$ ratio indicates
inward motion in the core. Line ratios similar to the SMM1 North spectrum were
predicted for a subset of their models.

In the case of SMM1 South the HCN(1-0) spectrum is detected
with a higher signal-to-noise ratio of up to 40.
By fitting the line wings with a symmetric profile, we get
v$_{\rm LSR}=34.8$\,km\,s$^{-1}$.
The hyperfine components
are characterised by complex line shapes and cannot be described
within the optically thin LTE approximation. The main features
of the line profiles are strong dips near the line centres.
In essence, such dips can be interpreted: 1) as an artifact of the data
combination, i.e.\ the short-spacing flux is not recovered when combining the
interferometric and single-dish data; 2) as a physical effect, namely,
as a result of the self-absorption of the internal radiation in the envelope.
To illustrate that this is not an artifact of data reduction we show
the single-dish and interferometric HCN(1-0) spectra
(integrated over $20\arcsec\times20\arcsec$)
before their combination in the right panels of Fig.~\ref{fig:pdbihcn}.
If the dips in the interferometric spectrum are
a result of the missing short-spacing flux, then we should see prominent emission
near the line centres in the single-dish data.
However, both interferometric and single-dish spectra have strong dips that
indicate that these line features are real.

%\clearpage
%\clearpage

\section{Discussion}

\subsection{The fragmentation and column density of the SMM1 clump}

Our observations reveal two major fragments towards the SMM1 clump.
From its mean FWHM extent of 30000\,AU \citep{2006ApJ...637..380B} we calculate
a volume-averaged density of n$_{\rm H_2} = 9\times10^4$\,cm$^{-3}$ for SMM1.
The dust temperature of 16.5\,K gives a Jeans length of about 17000\,AU and
a Jeans mass of 0.7\,M$_\sun$ \citep{2005fost.book.....S}.
This indicates that the two cores with masses more than one magnitude higher
are not the direct result of thermal
fragmentation, because one would then expect a number of cores with about one
Jeans mass to form \citep[e.g.][]{2005A&A...435..611J,2006MNRAS.368.1296B}.
The projected distance of the two cores is 21000\,AU.
It is rather large
compared to the radius of influence of radiative feedback from protostellar
objects. The simulations of \citet{2007ApJ...656..959K} show that the latter is
around 1000\,AU in the first 20000\,yr of core collapse.
This suggests that the two cores evolve individually in terms of radiative
feedback, however a kinematic influence is not excluded.
\citet{2008Natur.451.1082K} proposed a lower column density threshold of
1\,g\,cm$^{-2}$ (N$_{\rm H_2}=2\times10^{23}$\,cm$^{-2}$)
for the formation of high-mass stars.
For the SMM1 clump, we derived a peak column density of $7\times10^{22}$\,cm$^{-2}$
from the single-dish 1.2\,mm map peak flux.
It does not reach the proposed limit.
However, the core peak column densities lie beyond the threshold
(see Table~\ref{tab:corepars}). This shows that
observations with high spatial resolution are required to evaluate the
core properties.

\subsection{Properties of the two millimetre cores}
\label{sec:coreproperties}

The two detected cores SMM1 North and South appear quite similar at long
wavelengths because they are of nearly the same size and exhibit
comparable continuum emission. However, the derived core masses depend crucially
on the assigned dust temperatures, and these are determined by the
flux levels in the far-infrared. At wavelengths below $\sim$100\,$\micron$ the SMM1
clump of about 75\,M$_\sun$ is expected to become optically thick for continuum
emission, and our Planck component fit does not take this into account.
We therefore neglect the fitted warm component for SMM1 South in the rest of the
discussion. Also for the cold components, we may have underestimated the
characteristic dust temperature.
However, because of its mass the far-infrared continuum optical depth of SMM1 is
expected to be of the order of unity only. This means that a significantly
higher dust temperature, corresponding to a lower mass and in consequence to a
lower optical depth, would not be consistent with the far-infrared fluxes.
Thus, the resulting uncertainty of the masses are about a factor of two, but
the general averaging along the line-of-sight could introduce larger errors.
Furthermore, our mass derivation relies on the used dust opacities. While our
choice of the OH5 model is rather conservative and the results can be
compared to other studies, the true opacities may well differ. In the following,
we take the derived masses at face value.

Both SMM1 North and South are compact and appear to drive individual outflows.
Therefore we suspect them to form individual stars or stellar systems of a few.
With about 10 and 15\,M$_\sun$, SMM1 South and North represent cores of
intermediate mass.
From the comparison of core and stellar mass distributions,
there are tentative indications that the star-forming efficiency on the scales
of individual cores is between 25 and 50\%
\citep[see e.g.][and ref.\ therein]{2008ASPC..387..200K}.
Under this assumption, one would expect both cores to form intermediate-mass
protostars or protostellar systems in case none of the surrounding matter enters
in the core collapse. However, the latter is probable because of the
large-scale collapse motions that have been observed towards SMM1.
In comparison to samples of low-mass dense cores, SMM1 North and South are
denser by a factor of 10 \citep[cf.][]{2007A&A...476.1243M}.
We regard the peak column density values we derived from the
interferometric 1.3\,mm peak fluxes as lower
limits because of the filtering in the interferometric map.
For low-mass dense cores, peak column densities of $10^{23}$\,cm$^{-2}$ or lower
were reported \citep{1999MNRAS.305..143W}, so both cores lie above.
They rather resemble the objects found in IRDCs by
\citet{2007ApJ...662.1082R}. While their sample of four clumps
(we use a different nomenclature here) have masses of 100\,M$_\sun$ and greater,
the small scale cores are several thousand AU in size and their masses are
2 to 21\,M$_\sun$ with the exception of one hot molecular core candidate.
As in the case of SMM1 South, 24\,$\micron$ sources are associated with those cores.
Similar findings have been reported by \citet{2007ApJ...656L..85B} for the
more massive core in IRDC 18223-3, in particular the remarkable low luminosity.

The SED of SMM1 South at wavelengths below about 80\,$\micron$ (Fig.~\ref{fig:sed_both})
does not resemble the single thermal emission component reproducing the
submillimetre and millimetre fluxes. A similar result was also derived in the study
of a sample of 12 clumps detected in five other ISOSS star-forming regions
\citep{2008A&A...485..753H} and for one core in the ISOSS J23053+5953 region
\citep{2007A&A...474..883B}. Warm and hot dust components are required
to explain the emission down to 24\,$\micron$, and also emission from very small
grains may contribute in this wavelength regime. The fact that SMM1 South appears
as point source at 24 and 70\,$\micron$ shows that this core contains a compact region
of heated dust. Assuming half of the PSF FWHM (6$\arcsec$ at 24\,$\micron$) as an upper
limit, the
emitting region has a size of less than 7000\,AU. This is consistent with an
embedded young stellar precursor that constitutes the driving source
of the SMM1 South outflow.

SMM1 North lacks emission in the infrared. The only sign of star formation
is the SMM1 North outflow, indicating that this core is also further evolved
than a supposed prestellar stage. The characteristic dust temperature is
constrained to an upper limit, and therefore in this case the derived core mass
represents only a lower limit.

Regarding the motion of the two cores with respect to each other, their
systemic velocities derived from the HCN spectra differ by about
1.3\,km\,s$^{-1}$ and SMM1 North appears to be receding.
The estimated uncertainty is 0.5\,km\,s$^{-1}$.
Nevertheless, the value is relatively high compared to the velocity dispersions
of up to 0.5\,km\,s$^{-1}$ measured for the core-to-core motions e.g.\ in
NGC~1333 and Perseus \citep{2007ApJ...655..958W,2007ApJ...668.1042K}. It
corresponds to roughly 0.3\,AU\,yr$^{-1}$, and on a timescale of about
10$^5$\,yr the cores would cover their projected distance.

\subsection{The star-forming process within the cores}

The SEDs of both cores are dominated by the emission at long wavelengths
arising from cold dust. This emission does not disclose much information
about the internal structure of the cores. The lack of mid-infrared emission
and the low spatial resolution in the far-infrared, when compared to the expected
scales of emitting regions, do also prevent clarification of the properties of
the embedded source in the case of SMM1 South. SMM1 North is very likely less
evolved, but we cannot exclude projection effects.
More observations are needed to further constrain the star formation process in
the detected cores. The core emission morphology in the far-infrared will be
explored with the ESA Herschel mission. In the case of SMM1 South, high spatial
resolution imaging beyond 20\,$\micron$ combined with high sensitivity, possible
with the MIRI instrument on JWST, will severely constrain the properties of the
embedded protostellar object.
In the following, we
discuss implications given by the derived outflow properties and our modelling
of the HCN spectra.

\subsubsection{Outflow activity}

Collimated outflows have been observed for low-mass cores and, more recently,
also for young high-mass objects, supporting the disk accretion scenario
\citep[see][and ref.\ therein]{2005ccsf.conf..105B}.
In the former case, a high degree of
collimation is linked to early stages and a widening of the outflow opening
angle follows during further evolution \citep{2006ApJ...646.1070A}.
According to some observational evidence, this holds for high-mass sources
as well and may provide a basis for their evolutionary classification.

The fact that we observed rather collimated outflows close to their proposed
origins SMM1 North and South is consistent with the idea that the embedded
driving sources have formed and launched the outflows recently. This is
supported by the estimated dynamical timescales of less than $10^4$\,yr.

The outflow energetics change by orders of magnitude from low-mass cases
to luminous high-mass driving sources
\citep[e.g.][]{2004A&A...426..503W,2000A&A...353..211H}, and
the span of outflow kinetic energies that have been derived is
$10^{31}$\,J $\leq $ E$_{\rm kin} \leq 10^{41}$\,J.
For the sample of \citet{2002A&A...383..892B} with L $> 10^3$\,L$_\sun$, they
got values on the order of E$_{\rm kin} \approx 10^{39}$\,J.
With E$_{\rm kin} \approx 5\times10^{38}$\,J, the outflows of the SMM1 cores
reach the same order of magnitude, in particular because the adopted masses are
lower limits.
They exceed those of low-mass cores.

So far we have only considered the molecular outflows close to the cores.
The traced molecular gas has probably been entrained by collimated jets.
If we assume that the H$_2$ emission features at larger distances stem from
these jets, which is supported by the rough alignment with the molecular
outflows, we can derive a second timescale for the outflow activity of the
cores. However, we cannot constrain the jet velocities from observations.
\citet{2002A&A...383..892B} used a ratio of jet velocity to molecular outflow
velocity of 20, and for the SMM1 cores the resulting jet velocities are
about 300\,km\,s$^{-1}$. This is in the range of velocities that have been
observed for Herbig-Haro flows from low-luminosity sources \citep{2001ARA&A..39..403R}.
From the measured
offsets of the outermost H$_2$ feature in the south-west we got a timescale of
approximately $1.8\times10^4$\,yr for the SMM1 South jet. This is above the
molecular outflow timescale of approximately $7\times10^3$\,yr. For the SMM1
North jet, the outermost feature in the west gives a timescale of about
$7\times10^3$\,yr, which agrees with the molecular outflow timescale.
Accounting for the overall uncertainties, these values are consistent with the
proposed relation of jets and molecular flows, which of course does not exclude
other scenarios.

Parsec-scale jets and outflows have been found towards many young low-mass
stellar objects, mostly traced by optical or near-infrared line emission
\citep{2001ARA&A..39..403R}.
The more energetic outflows in the high-mass regime are therefore expected to
also extend to these sizes, although few observations were reported to date
\citep[e.g.][]{2008ASPC..387..158B,2002A&A...387..931B}.
The presumed jets from SMM1 support this idea.
The filamentary and patchy structure observed in H$_2$ can be caused by different
factors.
Besides the varying line-of-sight extinction, the local H$_2$ abundance may play
an important role in determining where the shock-excited emission arises, and
the kinematics of internal shocks is presumably influenced by the penetrated
medium.
An interesting possibility is the connection to the mass ejection history of the
driving source, which is linked to the accretion history for disk-driven
outflows.
The several features we detected indicate a varying mass-loss rate for the
SMM1 cores.
Such a burst mode of accretion may result from disk instability driven by infall
\citep{2006ApJ...650..956V}.

The S-shape that we observe for the presumed SMM1 South jet can be interpreted
as a precession of the outflow axis. The outermost feature we found
in the south-west approximately lies on the projected outflow axis with
PA $\approx 50\arcdeg$ we derived from the molecular outflow lobes close to the
core, while the features in between are offset to lower PA. This means the core
rotation axis is not along the line-of-sight, but rather has a high inclination.

\subsubsection{Modelling of the HCN emission}

HCN appears to be more resistant to freeze-out in cold dense cores compared to
CO-related molecules as HCO$^+$ \citep{2008ASPC..387...38R}.
Assuming that the HCN emission traces the dense gas in the cores,
we investigate in the following if the observed HCN(1-0) spectra can be,
in principle, reproduced by simple models for the core.
First, we considered the ``one-layer'' model where the core is spherically
symmetric and homogeneous.
We assumed that the observed spectrum is obtained towards the centre of the
core as the antenna beam size is little smaller than the spatial extent of the
core.
The parameters of the model are:
the hydrogen density n$_{\rm H_2}$, the temperature
T$_{\rm kin}$, the molecular column density N$_{\rm HCN}$, the turbulent
velocity v$_{\rm turb}$, and the regular velocity v$_{\rm rad}$
which characterises the radial expansion or contraction of the core.
Note that we do not specify the radius of the core as well as the actual
abundance of HCN since the emergent spectrum for such a core
depends on their product \citep[cf.][]{2008arXiv0808.2375P}.
Therefore, in the frame of this model we cannot independently constrain the
relative HCN abundance, the radius of the core, or its mass.

Having specified the above parameters, we generated the model core and performed
a line radiative transfer (LRT) simulation with the non-LTE code
of \citet{2004ARep...48..315P} using molecular line data from
\citet{2005A&A...432..369S}.
A synthetic spectrum of HCN(1-0) was calculated from the LRT simulation.
In principle, the systemic velocity v$_{\rm LSR}$ is an additional parameter.
We have assumed the values of 36.1\,km\,s$^{-1}$ (SMM1 North) and
34.8\,km\,s$^{-1}$ (SMM1 South) derived from the positions of the spectral
features.
The consistency between calculated and observed spectra was evaluated with
a $\chi^2$-criterion and took into account all the velocity channels of the
combined HCN(1-0) spectrum.

\begin{figure*}
\epsscale{1.15}
\plotone{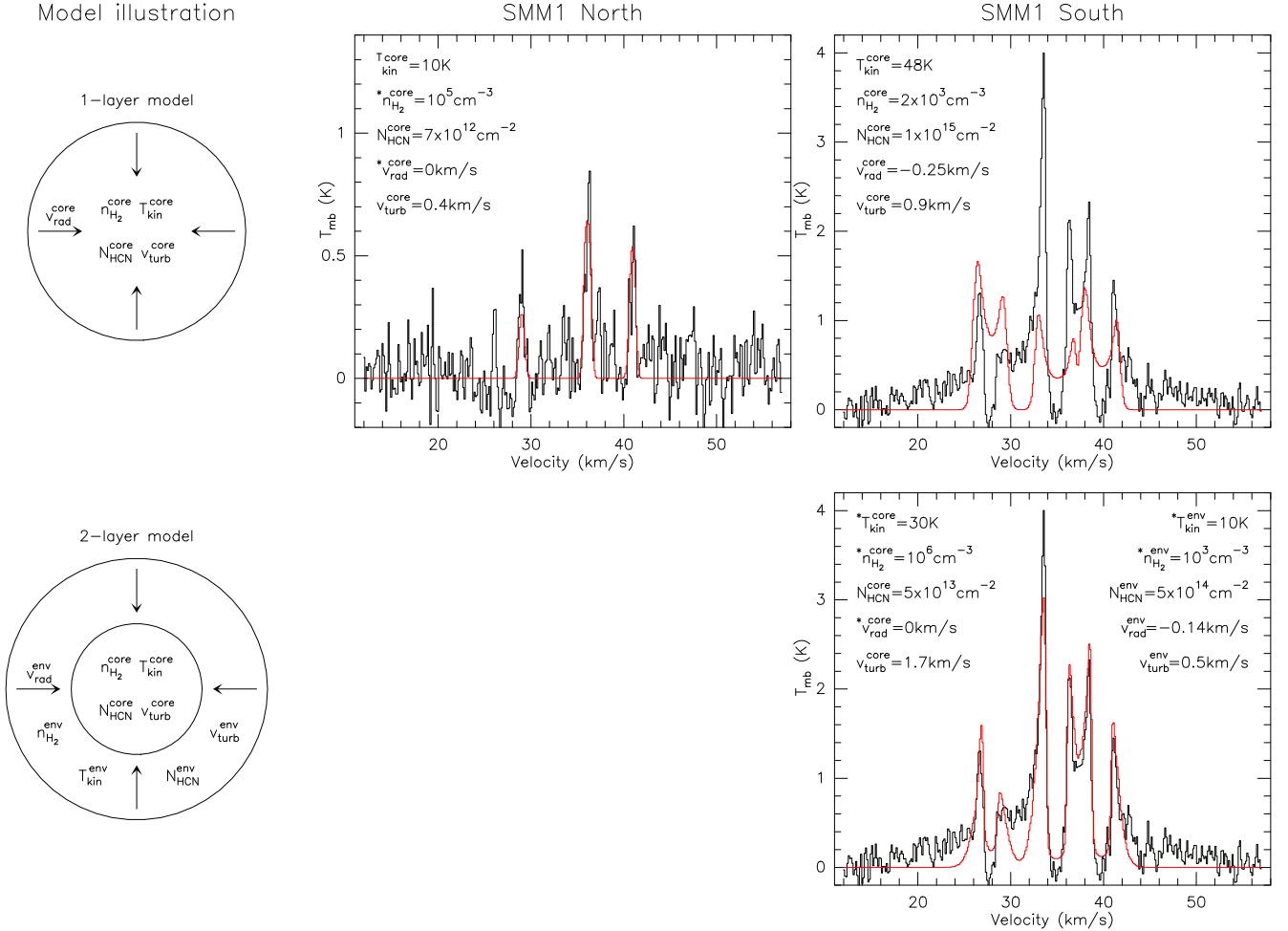}
\caption{Model illustration, observed and modelled HCN(1-0) spectra towards the
two SMM1 cores.
In the top row, the one-layer model is illustrated and the resulting best fit
spectra (red curves) are shown with the corresponding model parameters.
The bottom row shows the two-layer model and the resulting best fit of the
SMM1 South spectrum (red curve) as well as the corresponding model parameters.
Parameters that have been fixed are marked with a preceding asterisk (*).
}
\label{fig:hcnspec}
\end{figure*}

To search for the best set of the model parameters, we used Powell's minimisation
algorithm \citep{1992nrfa.book.....P}.
We had to specify parameter ranges for the search for the best-fit model.
In our simulations we used the ranges given in the Table~\ref{tab:fitpars} to
derive the best-fit parameter sets.
The ranges resulted from several runs where we specified parts of the adjacent
parameter space and did not find good reproductions of the observed spectra.
Though unlikely, we cannot exclude that additional compatible sets of
parameters exist.
In order to assure that the minimisation routine was not trapped in a local
minimum, we repeated the calculations starting with several sets of initial
parameters.
The best-fit spectra obtained for the one-layer model with the corresponding
parameters are shown in Fig.~\ref{fig:hcnspec} in the top row.

In the case of SMM1 North, the observed line ratios were reproduced fairly well.
We found a degeneracy between the hydrogen density, the kinetic temperature and
the HCN column density and therefore we set n$_{\rm H_2}=10^5$\,cm$^{-3}$.
This resulted in T$_{\rm kin} = 10$\,K and
N$_{\rm HCN}=7\times10^{12}$\,cm$^{-2}$.
Furthermore, given the assumption that the observed HCN(1-0) components are
indeed single-peaked and relatively narrow, the effect of the regular and
turbulent velocities cannot be distinguished.
Therefore we set the regular velocity to zero.
The derived value of the turbulent velocity is 0.4\,km\,s$^{-1}$.
Note that in contrast to the study of \citet{1993A&A...279..506G}, we did not
have to introduce any regular velocity to explain the ``anomalous'' line ratios.
The obtained ratios naturally appeared as a result of the non-LTE excitation in
the subcritical density.

In the case of SMM1 South, the one-layer model could not reproduce the maximal
intensities and the strength of the self-absorption dips at the same time.
We show one representative spectrum from the one-layer model in
Fig.~\ref{fig:hcnspec}.

A natural way to reproduce the observed spectrum of SMM1 South is to
include a low-density envelope in the model which should lead to self-absorption.
We checked this by constructing a ``two-layer'' model with 5+5 parameters,
and chose a configuration where a static core (v$_{\rm rad}^{\rm core}=0$) is
surrounded by a spherically symmetric, homogeneous envelope.
The increased number of parameters makes it difficult to identify a
best set through the minimisation routine.
Therefore we fixed some of the parameters based on the following arguments.
First, we assumed that the hydrogen
density in the core is higher while in the envelope it is lower than
the critical density for HCN(1-0).
Given this assumption, we expect the formation of strong self-absorption
dips if the HCN column density in the envelope is high enough.
We set the densities to n$_{\rm H_2}^{\rm core}=10^6$\,cm$^{-3}$ and
n$_{\rm H_2}^{\rm env}=10^3$\,cm$^{-3}$.
We also fixed the temperature in the envelope T$_{\rm kin}^{\rm env} = 10$\,K
assuming that it is not heated by the inner (most probably warmer) part of the
core, for which we chose T$_{\rm kin}^{\rm core} = 30$\,K.
Changes in the core temperature do not affect the derived model spectra significantly.
Thus, we varied the five parameters
N$_{\rm HCN}^{\rm core}$, v$_{\rm turb}^{\rm core}$, N$_{\rm HCN}^{\rm env}$,
v$_{\rm turb}^{\rm env}$, v$_{\rm rad}^{\rm env}$, and constrained them using
the minimisation routine.

The best-fit spectrum obtained from the two-layer model is presented
in Fig.~\ref{fig:hcnspec} (bottom).
The two-layer model could reproduce the intensity, the strong self-absorption
dips and asymmetry of the observed spectrum quite well.
The line wings may stem from outflowing gas, or also be due to a more complex
velocity profile not incorporated in the model.
The important outcome of this model can be summarised as follows:
First, the strong self-absorption in the line profiles can be reproduced by the
high-density core together with the low-density envelope.
Second, the asymmetry of the line profiles can be explained by the infalling
envelope (we got v$_{\rm rad}^{\rm env}=-0.14$\,km\,s$^{-1}$),
given the relatively high turbulent velocities in the core
(v$_{\rm turb}^{\rm core}=1.7$\,km\,s$^{-1}$)
and in the envelope (v$_{\rm turb}^{\rm env}=0.5$\,km\,s$^{-1}$).
However, this configuration is not the only way to fit the line asymmetries.
In particular, we were able to reproduce the line asymmetries by the
combination of an expanding core and a static envelope because of the similar
relative velocities.
We consider this scenario implausible, though, because of the short outflow
timescales.

Finally, we stress that the considered simple models did not reliably constrain
the other parameters of the cores, but they illustrated that the observed
spectra can indeed be explained theoretically.
In order to get more reliable information about the core properties
one has to consider a set of molecular lines and transitions together with
more physical models of the source.
Unfortunately, the available data, in particular the HCO$^+$(1-0)
measurements, are not sufficient to enable this.
Indeed, if we used HCO$^+$(1-0) in addition to HCN(1-0) in our fitting routine,
we would have to incorporate two more parameters for the HCO$^+$ abundance
(in the core and in the envelope), while HCO$^+$(1-0) alone will not allow us
to further constrain the excitation conditions.
For the latter, at least a few more transitions of HCO$^+$ or its isotopomers
were necessary.
On the other hand, a more physical model of the core (e.g.\ with smooth
distributions) has more parameters to fit which makes such a model even
more uncertain.
Thus, the usage of more sophisticated models is required when detailed spectral
maps of the sources are available.

%\clearpage
\begin{deluxetable*}{lcc}
\tablecaption{Parameter ranges used in the HCN(1-0) emission modelling for the
one-layer and two-layer models.
\label{tab:fitpars}}
\tablehead{
\multicolumn{3}{c}{One-layer Model}\\
\colhead{Parameter}
& \colhead{SMM1 North}
& \colhead{SMM1 South}
}
\startdata
n$_{\rm H_2}$ (cm$^{-3}$)       & $10^{5}$                & $10^{3}\dots10^{5}$ \\
T$_{\rm kin}$ (K)               & $10\dots50$             & $10\dots50$         \\
N$_{\rm HCN}$ (cm$^{-2}$)       & $10^{12}\dots10^{15}$   & $10^{13}\dots10^{15}$ \\
v$_{\rm turb}$ (km\,s$^{-1}$)   & $0\dots1.0$             & $0\dots1.0$ \\
v$_{\rm rad}$ (km\,s$^{-1}$)    & $0$                     & $-0.25\dots0$ \\
\hline
\multicolumn{3}{c}{Two-layer Model for SMM1 South\rule{0pt}{20pt} }\\
Parameter & Core & Envelope\\[5pt]
\hline
n$_{\rm H_2}$ (cm$^{-3}$)       & $10^{6}$                & $10^{3}$\rule{0pt}{15pt} \\
T$_{\rm kin}$ (K)               & $30$                    & $10$    \\
N$_{\rm HCN}$ (cm$^{-2}$)       & $10^{13}\dots10^{15}$   & $10^{13}\dots10^{15}$ \\
v$_{\rm turb}$ (km\,s$^{-1}$)   & $0\dots3.0$             & $0\dots1.0$ \\
v$_{\rm rad}$ (km\,s$^{-1}$)    & $0$                     & $-1\dots1$ \\
\enddata
\end{deluxetable*}
%\clearpage

%\clearpage

\subsubsection{Collapse of SMM1 South}

The modelling of the HCN emission showed that the spectrum of SMM1 South can be
explained with a collapse of the core.
From the resulting velocity we calculated estimates for the mass infall
rate using dM/dt $\approx$ M/t $ = $ M\,v$_{\rm in}$/r.
For SMM1 South, v$_{\rm in} = 0.14$\,km\,s$^{-1}$ is the regular velocity of the
envelope in the two-layer model. Together with the mass and radius from
Table~\ref{tab:corepars} this gave an infall rate of
$3\times10^{-5}$\,M$_\sun$\,yr$^{-1}$.
This is consistent with the mass outflow rates, but is only a crude estimate.

\subsubsection{Internal turbulence}

Additional parameters we obtained from the HCN emission are the internal
turbulent velocities of the cores that reproduced the observed line widths.
However, in both cases rotational motions, more complex velocity profiles,
and also outflows
may contribute to the line widths. Such effects were not considered in our
models.
In the case of SMM1 South, the fitted model shows a very turbulent core and
less turbulence in the envelope. When compared to the isothermal sound speed of
around 0.2\,km\,s$^{-1}$ for cold cores (T $\approx 20$\,K), the core
turbulence of 1.7\,km\,s$^{-1}$ is highly supersonic. The turbulent velocity in
the envelope of 0.5\,km\,s$^{-1}$ is on the scale of the sound speed but still
in the supersonic regime.
The SMM1 North HCN spectrum indicates a more quiescent state of this core.
The linewidths are reproduced with a turbulent velocity of 0.4\,km\,s$^{-1}$.

\section{Summary and conclusions}

\begin{figure}
\epsscale{1.15}
\plotone{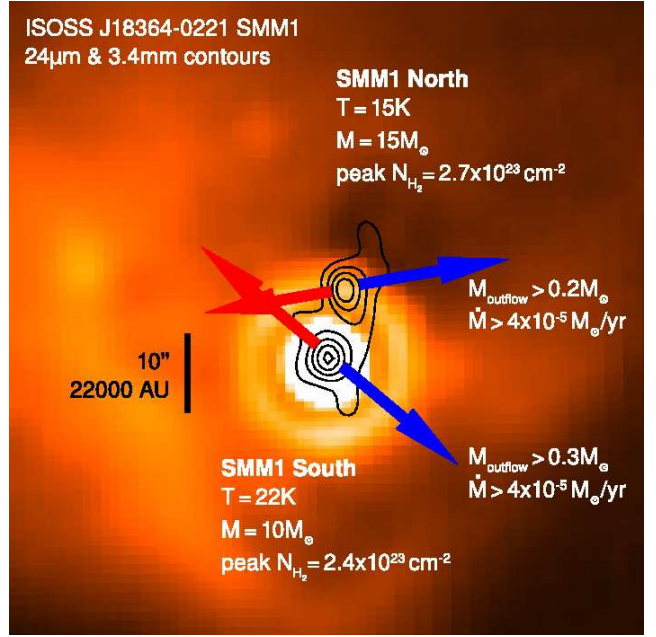}
\caption{Schematic illustration of the identified cores, the corresponding
outflows, and their properties.
The 3.4\,mm contours are plotted above the 24\,$\micron$ emission
(cf.\ Fig.~\ref{fig:4map}).
}
\label{fig:schem}
\end{figure}

We have investigated the onset of star formation in the massive cold clump
\object{ISOSS J18364-0221} SMM1.
The derived results regarding the clump fragmentation, the core and outflow
properties, and the collapse indications are summarises as follows:
\begin{enumerate}
\item Two compact, embedded cores are the only objects we identified.
Thus, this clump shows little fragmentation, but seems to rather produce a low
number of compact objects.
\item The cores dubbed SMM1 North and South are of about 10000 and 9000\,AU radius.
The dust temperatures are 15 and 22\,K and
they have masses of about 15 and 10\,M$_\sun$, respectively.
SMM1 South harbours an infrared source not detected at 8, but at 24 and 70\,$\micron$.
The luminosity of SMM1 South is about 180\,L$_\sun$ and a molecular outflow
is detected in its vicinity.
SMM1 North lacks infrared emission, but a second molecular outflow is emerging
from this core. The luminosity of SMM1 North is approximately 20\,L$_\sun$.
Both outflows appear collimated and we derived lower outflow mass limits of
about 0.2 and 0.3\,M$_\sun$ of presumably entrained gas, and mass outflow rates of
about $4\times10^{-5}$\,M$_\sun$\,yr$^{-1}$.
In Fig.~\ref{fig:schem} we show a schematic illustration of these results.
\item We detected filaments of shock-excited H$_2$ emission at 2.122\,$\micron$ towards
the lobes of the SMM1 South molecular outflow, and also at distances up to
4.6\,pc from SMM1. Each filament is roughly aligned with one of the outflow axes.
They are possibly tracing the outflow-generating jets, and the dynamical
timescales we derived are consistent with those of the molecular outflows
considering the uncertain jet velocities.
For both molecular outflows, ages of less than $10^4$\,yr were found.
\item SMM1 South is bright in HCN(1-0) and our modelling results support that the core
is collapsing. We got an infall velocity of 0.14\,km\,s$^{-1}$. This results in
a mass infall rate estimate of $3\times10^{-5}$\,M$_\sun$\,yr$^{-1}$.
For both cores the spectra can be interpreted with turbulent internal motions,
and in the case of SMM1 South the innermost part appears highly supersonic.
\end{enumerate}

Both SMM1 North and South are more massive than typical low-mass dense cores
and embedded in the presumably contracting 75\,M$_\sun$ SMM1 clump.
This suggests that the two cores harbour protostellar seeds that
may become intermediate- to high-mass stars.
The presence of a 24\,$\micron$ source, a very turbulent central region
and the jet features at large distances to the core support that SMM1 South is
more evolved than SMM1 North. We interpret it as having an embedded young
protostar that constitutes the driving source of the outflow.
The outflows as well as the infall indicate that it is accreting.
The associated outflow indicates that SMM1 North is also a star-forming core,
and the outflow energetics are similar. However, the forming object remains
undetected, and the relatively low level of turbulence implies that it has only
marginally affected the surrounding core. The presumed difference in evolution
of the forming objects contrast with the similar sizes, densities and outflow
properties of the cores. These findings suggest a rapid evolution of the
luminosity as predicted for high accretion rates \citep{Yorke08}.

In comparison to the core collapse simulations of \citet{2007ApJ...656..959K} we
find that the luminosity of SMM1 South would be reached at a stage where the
embedded protostellar mass is around 0.3\,M$_\sun$. Accretion
onto the protostar is the dominant luminosity source, and they find an
accretion rate on the order of 10$^{-4}$\,M$_\sun$\,yr$^{-1}$ from directly
after the formation of the protostar until this stage. The resulting timescale
of about 3000\,yr is consistent with the outflow timescales.
Also the mass outflow rate as well as the infall rate estimate may
be consistent with the high accretion rate.
SMM1 North is less luminous, and hence the protostellar mass would be even
lower, but may still be consistent with a high accretion rate in their model.
However, the formation of outflows is not incorporated therein.

%\clearpage
%\clearpage

\acknowledgments

We thank the referee for the helpful comments to improve this manuscript.
We also thank J\"urgen Steinacker, Henrik Beuther and Ulrich Klaas (MPIA) for their
support and helpful discussions; Jan Martin Winters, Pierre Hily-Blant and
Roberto Neri (IRAM) for their help reducing the IRAM data.
We acknowledge the support by DLR (German Space Agency) through grant 50 OS 0501.
MH is fellow of the International Max Planck Research School for Astronomy and
Cosmic Physics at the University of Heidelberg (IMPRS-HD).
Based on results of ISO, an ESA project with instruments funded by ESA
Member States (especially the PI countries: France, Germany, the Netherlands and
the United Kingdom) and with participation of ISAS and NASA. The ISOSS was
supported by funds from DLR.
Based on observations collected at the German-Spanish Astronomical Centre at
Calar Alto, operated jointly by the Max-Planck-Institut f\"ur
Astronomie and the Instituto de Astrof\' isica de Andaluc\'ia (CSIC),
on observations with the James-Clerk-Maxwell Telescope (JCMT),
on observations made with the Spitzer Space Telescope, which is operated by the
Jet Propulsion Laboratory, California Institute of Technology under a contract
with NASA,
and on observations at the IRAM 30m and the IRAM Plateau de Bure Interferometer.
IRAM is supported by CNRS (France), MPG (Germany), and IGN (Spain).
This publication makes use of data products from the Two Micron All Sky Survey,
which is a joint project of the University of Massachusetts and the
IPAC/California Institute of Technology, funded by the
NASA and the NSF, and made use of NASA's ADS Bibliographic Services.

{\it Facilities:} \facility{CAO:3.5m (Omega-2000)},
\facility{Spitzer (IRAC, MIPS)}, \facility{JCMT (SCUBA)},
\facility{IRAM:30m (VESPA, MAMBO-2)}, \facility{IRAM:Interferometer}

%% The reference list follows the main body and any appendices.
%% Use LaTeX's thebibliography environment to mark up your reference list.
%% Note \begin{thebibliography} is followed by an empty set of
%% curly braces.  If you forget this, LaTeX will generate the error
%% "Perhaps a missing \item?".
%%
%% thebibliography produces citations in the text using \bibitem-\cite
%% cross-referencing. Each reference is preceded by a
%% \bibitem command that defines in curly braces the KEY that corresponds
%% to the KEY in the \cite commands (see the first section above).
%% Make sure that you provide a unique KEY for every \bibitem or else the
%% paper will not LaTeX. The square brackets should contain
%% the citation text that LaTeX will insert in
%% place of the \cite commands.

%% We have used macros to produce journal name abbreviations.
%% AASTeX provides a number of these for the more frequently-cited journals.
%% See the Author Guide for a list of them.

%% Note that the style of the \bibitem labels (in []) is slightly
%% different from previous examples.  The natbib system solves a host
%% of citation expression problems, but it is necessary to clearly
%% delimit the year from the author name used in the citation.
%% See the natbib documentation for more details and options.

\end{document}